\documentclass[prl, aps, superbib, tightenlines, twocolumn, floatfix, showpacs,amssymb]{revtex4}
\usepackage{amsmath, graphicx, dcolumn, bm}

\renewcommand{\vec}{\bf}
\begin{document}

\title{Turbulence-induced magnetic flux asymmetry at nanoscale junctions}
\author{Neil Bushong}
\email{bushong@physics.ucsd.edu}
\affiliation{
Department of Physics, University of California, San Diego, La Jolla,
CA 92093-0319}

\author{Yuriy Pershin}
\email{pershin@physics.ucsd.edu}
\affiliation{
Department of Physics, University of California, San Diego, La Jolla,
CA 92093-0319}

\author{Massimiliano {Di Ventra}}
\email{diventra@physics.ucsd.edu}
\affiliation{
Department of Physics, University of California, San Diego, La Jolla,
CA 92093-0319}

\date{\today}

\begin{abstract}
It was recently predicted [J. Phys.: Condens. Matter {\bf 18}, 11059
(2006)] that turbulence of the electron flow may develop at
nonadiabatic nanoscale junctions under appropriate conditions. Here
we show that such an effect leads to an asymmetric current-induced
magnetic field on the two sides of an otherwise symmetric junction.
We propose that by measuring the fluxes ensuing from these fields
across two surfaces placed at the two sides of the junction would provide
direct and noninvasive evidence of the transition from laminar to
turbulent electron flow. The flux asymmetry is predicted to first
increase, reach a maximum and then decrease with increasing current,
i.e. with increasing amount of turbulence.
\end{abstract}


\maketitle

The hydrodynamics of the electron liquid dates back to earlier
studies by Madelung, Bloch~\cite{Madelung,Bloch} and later on by
Martin and Schwinger~\cite{MartinSchwinger}. In this latter work,
in particular, it was shown that the many-body time-dependent
Schr\"odinger equation (TDSE) can be written exactly in
hydrodynamic form in terms of the density $n({\bf r},t)$ and
velocity field ${\bf v}({\bf r},t)={\bf j}({\bf r},t)/n({\bf
r},t)$, the ratio of the current and charge density, with all
many-body interactions lumped into a two-particle stress tensor.

In recent years, the analogy of the electron flow with classical
fluid dynamics has been pushed even further with the development of
time-dependent density-functional methods and the consequent
realization that under certain conditions, the exchange-correlation
potentials can be written in hydrodynamic
form~\cite{vignale97,tokatly}. More recently, it was shown that
electron flow in nanoscale constrictions satisfies the conditions to
write the two-particle stress tensor in a form similar to the stress
tensor of the Navier-Stokes equations with an effective viscosity of
the electron liquid (see also below)~\cite{dagosta06jpcm}. The most
striking prediction of this result is that, under specific
conditions on the current, density and junction geometry, the
electron flow should undergo a transition from laminar to turbulent
regimes.~\cite{dagosta06jpcm} Recently, this behavior was confirmed
numerically by solving directly the TDSE within time-dependent
current-density functional theory~\cite{bushong07} and comparing the
results with the generalized Navier-Stokes equations derived in
Ref.~\cite{dagosta06jpcm}. In experiments, however, detecting
turbulence via direct imaging of the current density remains
challenging. For instance, scanning-probe microscopy (SPM)
experiments which image the current flow in a 2D electron gas (2DEG)
have been reported~\cite{topinka00}. These experiments employ an SPM
tip to reflect electrons back toward the junction, and measure the
resultant change in the total current. This means that the image
thus obtained gives the correlation between the tip position and
junction current, which does not necessarily correspond to the
magnitude of the current density. Moreover, SPM-type experiments
necessarily disturb the electron flow at the tip position, and are
therefore essentially invasive.

Another way to probe turbulence would be to measure the noise
properties of the current.  However, the analytical description of
turbulence is a notoriously intractable problem, thus making it
unclear what noise characteristics the turbulent flow of electrons
would generate. In addition, this noise would necessarily correlate
with other intrinsic types of noise, especially shot
noise~\cite{chediventra2003}.

In the present Letter we show that the measurement of the
current-induced magnetic field at the two sides of an otherwise
symmetric nanojunction provides a direct and non-invasive way of
measuring the transition from laminar to turbulent flow. In
particular, we predict that the fluxes ensuing from the
current-induced magnetic field across two surfaces on the two sides
of the junction would at first become increasingly different with
increasing current. This asymmetry reaches a maximum, and then
decreases with further increase of the current. The measurement of these
fluxes is within reach of present experimental capabilities, and thus the observation and study of
this phenomenon would provide valuable insight into the transport properties of
nanoscale systems.

\begin{figure}
\includegraphics[width=3.4in]{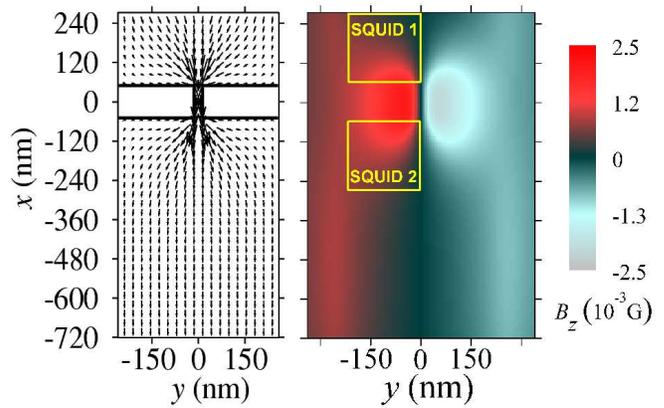}
\caption{(Color online) Laminar current flow at a low value of the
total current ($0.1 \, {\rm \mu A}$). Left panel: electron velocity
distribution. The arrow length is proportional to the velocity
magnitude. Right panel: normal component of the magnetic field
through a plane 50 nm above the 2DEG. The surface areas indicated
with SQUID 1 and SQUID 2 represent two areas across which we
calculate the magnetic flux (see text).} \label{fig:lamin}
\end{figure}

\begin{figure}
\includegraphics[width=3.4in]{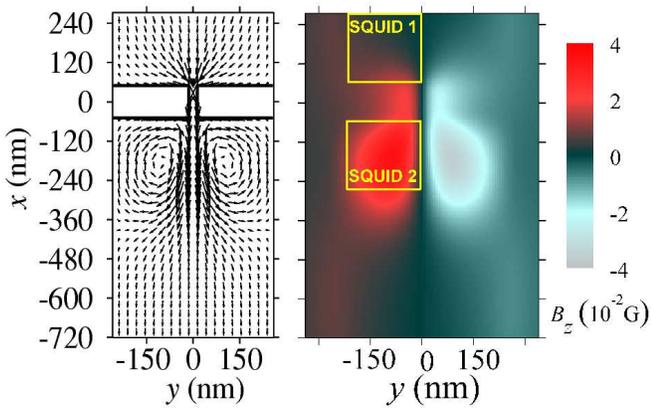}
\caption{(Color online) Turbulent current flow at high value of the
total current ($1 \, {\rm \mu A}$). Left panel: electron velocity
distribution. Right panel: normal component of the magnetic field
through a plane 50 nm above 2DEG. The top-down symmetry in magnetic
field distributions across the SQUID areas is lost. The magnetic
field flux through SQUID 2 is significantly higher then the flux
through SQUID 1.} \label{fig:turbul}
\end{figure}

The structure we have in mind consists of two symmetric regions of
a 2DEG connected {\em non-adiabatically} by a nanojunction (the
edges of this structure are represented with solid lines in the
left panel of Figs.~\ref{fig:lamin} and~\ref{fig:turbul}). The
non-adiabaticity requirement  is due to the fact that, as shown in
Ref.~\cite{dagosta06jpcm}, an adiabatic constriction produces a
Poiseuille flow, which is laminar for essentially all currents one
can effectively inject in a 2DEG.~\footnote{Note that if
turbulence is observed in a 2D system at a given current,
turbulence is even more favored in 3D at the same current.} The
lateral ($y$-direction) boundaries are closed to current flow, and
the longitudinal ($x$-direction) boundaries are open, with current
being injected in the ``top'' boundary and exiting in the
``bottom'' boundary. We then envision two identical surfaces --
placed at a given distance from the 2DEG in the $z$ direction --
across which we calculate the current-induced magnetic flux (see
Fig.~\ref{fig:lamin}). These magnetic fluxes can be measured by
two superconducting interference devices
(SQUIDs)~\cite{gallopsquids} located on the two sides of the
junction as illustrated in Fig.~\ref{fig:lamin}.

Our starting point is the time-dependent Schr{\"o}dinger equation
written in the approximate Navier-Stokes form for an incompressible
fluid~\cite{dagosta06jpcm}
\begin{gather}
D_t n({\vec r},t) = 0, \,\,
{\vec \nabla} \cdot {\vec v}({\vec r}, t) = 0, \nonumber \\
\begin{split}
m^* n({\vec r}, t) D_t v_i({\vec r}, t)
    = & -\frac{\partial}{\partial r_i} P({\vec r}, t)
    + \eta \nabla^2 v_i({\vec r}, t) \\
    & - n({\vec r}, t) \frac{\partial}{\partial r_i}
        V_{\rm ext}({\vec r}, t)
\end{split}
\label{eq:ns}
\end{gather}
where $D_t = \frac{\partial}{\partial t} + ({\vec v} \cdot {\vec
\nabla})$ is the convective derivative, $m^*$ is the effective mass
of the electrons, $n({\vec r}, t)$ is the electron density, $P({\vec
r}, t)$ is the pressure of the electron liquid, and $V_{\rm
ext}({\vec r}, t)$ is the sum of the Hartree and the ionic
potentials. The quantity $\eta=\hbar n f(n)$ is the viscosity of the
electron liquid, with $f(n)$ a smooth function of the density. The
values of the viscosity as a function of density have been
calculated using linear-response theory~\cite{conti99,Robert}; here,
we use the 2D interpolation formula of Ref.~\onlinecite{conti99}. We also employ the jellium 
approximation for the electron liquid, which together with the assumption of incompressibility, allows us 
to neglect spatial variations of $V_{\rm
ext}({\vec r}, t)$.
Incompressibility of the electron liquid represents to a good
approximation the behavior of metallic quantum point
contacts~\cite{dagosta06jpcm,bushong07}\footnote{Incompressibility
neglects the formation of surface charges near the edge of the gap
between the contacts~\cite{diventra02, sai07, bushong07}.  These
charges form dynamically during the initial transient of the
current, and during that time create a displacement current. The
latter would affect the initial-time magnetic field. However, this
surface charge distribution is stationary after the transient, and
therefore it does not influence the long-time behavior of the
magnetic field which is of interest here.}.

The current density was calculated numerically as a solution of
Eqs.~(\ref{eq:ns})~\cite{gerris} for a nanojunction 28 nm wide. We
have used Dirichlet boundary conditions for the velocity at the inlet,
and Neumann boundary conditions at the outlet. We use parameters
corresponding to a GaAs-based 2DEG: $m^* = 0.067m_{\rm e}$, and $n =
5.13 \times 10^{11} \, {\rm cm}^{-2}$. The calculations were performed
at a fixed value of the total average current flowing through the
system selected in the range from $0.001\, {\rm \mu A}$ to $10\, {\rm
\mu A}$. The magnetic field profile was found for each calculated
current density distribution.  The size of the surface area across
which we calculated the magnetic flux was chosen to be 200 $\times$
200 (nm)$^2$. Each surface is displaced laterally to one side of the
nanojunction as shown in Fig.~\ref{fig:lamin}. This surface represents
the SQUID area and is within reach of current
technology~\cite{lam03}, although, in principle, the SQUID's area does
not necessarily need to be so small. We assume that the surfaces are
located 50 nm above the 2DEG; therefore, the distributions of magnetic
field were calculated at this distance from the 2DEG. Increasing this
distance by a factor of two decreases the flux by about a factor of
two.~\footnote{An estimate of the flux decrease with distance can be
made considering the magnetic field flux -- through a square SQUID of
side $l$ -- created by a current in a wire located at a distance $h$
beneath a SQUID side. It can be shown that the magnetic field flux is
proportional to $\textnormal{Log}(1+h^2/l^2)$. Taking $l=200$ nm,
$h_1=50$ nm and $h_2=100$ nm we obtain the flux ratio of 1.76.} For
convenience, from now on we shall call these two surfaces, SQUID 1
area and SQUID 2 area (Figs.~\ref{fig:lamin} and~\ref{fig:turbul}).

The magnetic field fluxes allow us to characterize the degree of
asymmetry in the current flow pattern as well as to probe such
specific features of turbulent current flow as eddies. We find that
at low currents, the electron flow pattern is symmetric (see left panel of
Fig~\ref{fig:lamin}). More precisely, the patterns of current flow
in the two electrodes are mirror images of each other, with the
overall sign of the flow reversed~\cite{bushong07}. This is
illustrated in the left panel of Fig.~\ref{fig:lamin}, where we plot
the velocity distribution of the electron liquid at a simulation
time of 24 ps.  The total current is $0.1 \, {\rm \mu A}$, which is
small enough that the flow is in the laminar regime, as it is
evident from the figure. In the right panel of Fig.~\ref{fig:lamin},
we show the $z$-component of the magnetic field through a plane 50
nm above the 2DEG.  As is typical of the laminar regime, the
current-induced magnetic fields above the top and bottom contacts
are almost symmetric with respect to the center of the junction
producing an almost symmetric flux across the areas. Hence, SQUIDs
positioned as indicated in Fig. \ref{fig:lamin} would measure almost
equal magnetic fluxes.

By increasing the current, the current density in the source and
drain sides loses top-down symmetry: the current density in the
outgoing side becomes turbulent, while the current density in the
incident side remains laminar (Fig~\ref{fig:turbul}). We note that
at large currents we observe the formation of turbulent ``eddies''
which evolve in time, rather than a completely chaotic current
density distribution. This means that at the current values we
consider here turbulence is not fully developed~\cite{bushong07}.

Fig. \ref{fig:turbul} illustrates the behavior typical for the
turbulent regime. This plot corresponds to a total current of $1.0
\, {\rm \mu A}$ at time $t=24$ ps. As before, the left panel
illustrates the electron velocity distribution. Unlike the electron
velocity field presented in Fig. \ref{fig:lamin}, the
electron velocity distributions in the top and bottom electrodes in
Fig. \ref{fig:turbul} are no longer symmetric. In particular, the
electron velocity distribution in the bottom contact shows eddies
and an increased current density in the middle of the junction. Such
a velocity field is responsible
for a much stronger magnetic field above the bottom contact (in
particular, in the SQUID 2 area in Fig. \ref{fig:turbul}). The
magnetic field distribution above the top contact in
Fig.~\ref{fig:turbul} is ``smooth'' and uniform, and has a
structure similar to the magnetic field distribution above the top contact
of Fig. \ref{fig:lamin}.

\begin{figure}[b]
\includegraphics[width=3.4in]{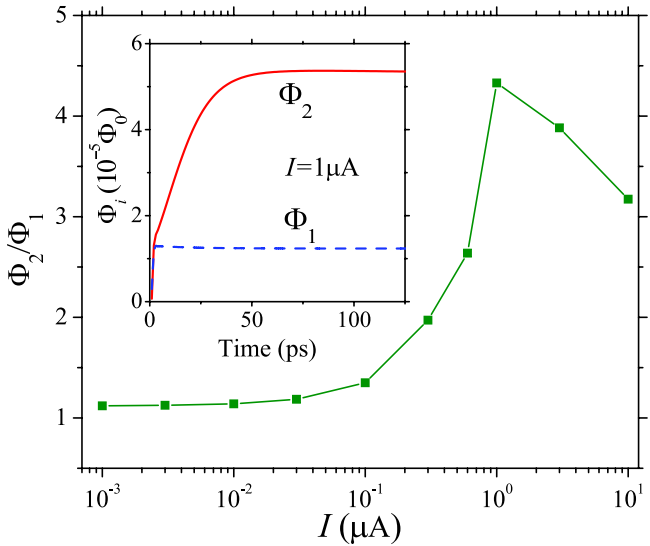}
\caption{(Color online) Asymptotic value of the ratio of the flux
through SQUID 2 area to the flux through the SQUID 1 area as a
function of the total current. See text for details. \emph{Inset}:
Time dependence of fluxes (in units of $\Phi_0=h/2e$) through the top (dashed blue line) and the
bottom (solid red line) SQUID areas, for a total current of $1.0 \,
{\rm \mu A}$.} \label{fig:asymp}
\end{figure}

For both the laminar and turbulent cases, as time passes, the fluxes
through the top and bottom SQUID areas saturate to constant values.
This is shown in the inset of Fig. \ref{fig:asymp} where we plot the
flux through the top and bottom SQUID areas as a function of time
for the $1.0 \, {\rm \mu A}$ case. We can now determine the
magnetic signature of the transition between the laminar and
turbulent regimes by plotting the asymptotic value of the ratio of
the magnetic fluxes through the two SQUID areas as a function of the
total current. This is illustrated in Fig. \ref{fig:asymp} and shows
the main findings of this paper.

At low currents (laminar regime), the current density in the top and
bottom contacts is highly symmetric. Therefore, the ratio of the
fluxes is near unity~\footnote{The flux ratio is expected to
approach unity in the limit of zero current and infinite systems.}.
There is a definite transition at the critical current $I_c
\backsimeq 0.3 \, {\rm \mu A}$, after which the ratio of the
magnetic fluxes increases by more than 300\%. The flux ratio reaches
a maximum at a current of about $1\mu$A. Increasing the current
further, we find that the flux ratio decreases. This can be
explained as follows. Near the critical current, the eddies are
small and stationary. These smaller, localized eddies lead to a
higher ratio $\Phi_2/\Phi_1$ and the increase we observe. On the
other hand, by increasing the current further, the eddies grow in
size and ``spread out'' away from the junction thus reducing the
flux ratio. Note that \emph{fully-developed} turbulence would
decrease the ratio $\Phi_2/\Phi_1$ even further, since eddies having
clockwise and counter-clockwise senses would be continuously generated
and destroyed. In addition, the noise level of modern
SQUIDs~\cite{noiseSQUID} is below $10^{-6}\Phi_0$ ( $\Phi_0=h/2e$ is the
magnetic flux quantum), which is well below the flux magnitude shown
in the inset of Fig. \ref{fig:asymp}.

\begin{figure}
\includegraphics[width=3.4in]{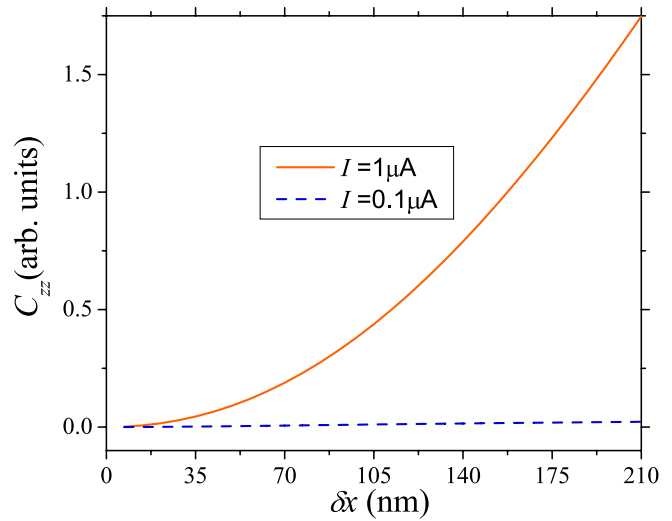}
\caption{(Color online)  Magnetic field correlation function, for
total currents of $1.0 \, {\rm \mu A}$ (solid orange curve) and
$0.1 \, {\rm \mu A}$ (dashed blue curve).  Because the total
current differs between the two by a factor of 10, we have scaled
the $0.1 \, {\rm \mu A}$ curve by a factor of $10^2$. Even after
this scaling, $C_{zz}$ is significantly larger in the turbulent
(high-current) case than it is in the laminar (low-current) case.}
\label{fig:magcorr}
\end{figure}

We conclude by quantifying the degree of turbulence of the current-induced magnetic field. While this cannot be directly measured,
it provides insight into the properties of the turbulent regime attainable experimentally. Let us then
calculate the magnetic field correlation tensor, which quantifies the spatial correlation of the magnetic field at different points in space.
We define this tensor as
\begin{equation}
C_{ij} = \langle (B_i({\vec r}) - B_i({\vec r} + \delta{\vec r}))
    (B_j({\vec r}) - B_j({\vec r} + \delta{\vec r})) \rangle \, .
\end{equation}
Here, $B_i$ and $B_j$ denote components of the magnetic field, and
$\delta{\vec r}$ is a given vector.  The brackets $\langle \ldots
\rangle$ denote averaging over all pairs of positions separated by
$\delta {\vec r}$ within a given region.  Note that, even before
performing a spatial average, the magnetic field already has a
nonlocal character, in that the magnetic field at a point is due to
the velocity of charges in the whole system.

In Fig. \ref{fig:magcorr} we plot the magnetic field correlation
tensor $C_{\rm zz}$, for $\delta {\vec r} = (\delta x, 0)$ at
$50$ nm above the 2DEG as a function of $\delta x$. The spatial
averaging was carried out in a region ``downstream'' from the
junction, in the region $x = [-487 \, {\rm nm}, -235 \, {\rm
nm}]$, $y = [0 \, {\rm nm}, 259 \, {\rm nm}]$.  As expected, in
the laminar case, the magnetic field varies with distance by a
small amount, so that $C_{zz}$ is small.  In the turbulent case,
the presence of the eddies leads to a magnetic field that
correlates spatially, causing $C_{zz}$ to increase with distance.

Finally, we discuss some possible alternatives to measuring turbulence
via the proposed magnetic fluxes. For one, instead of using two SQUIDs
placed at the two sides of the junction one could envision the use of
only one SQUID, by changing the direction of current flow by merely
reversing the bias. This also ensures that the effect of unavoidable
scattering by defects/impurities on the magnetic fluxes is accounted
for identically for both possible directions of overall current
flow. Another alternative is to use a movable
SQUID~\cite{scanningSQUID} or scanning Hall-probe
microscopy~\cite{Moler}. In particular, if an X-Y stage were added to
the apparatus, so that the substrate (or the SQUID) were movable, one
could generate images of the magnetic flux as a function of
position. For instance, this scanning SQUID microscopy has been
previously employed to study the spatial configuration of vortices in
Type II superconductors~\cite{veauvy04}. Imaging electron flow in this
way would provide information about the electric current density
throughout the device with the added benefit that the measurement
would be noninvasive~\footnote{Extracting the current density from the
magnetic field requires solving the ``inverse problem.''  This is
nontrivial, but the problem has been heavily studied; for example, the
practitoners of Magnetoencephalography (MEG)~\cite{liu02} have treated
the inverse problem in the context of medicine in order to image
electrical activity in the brain.}. However, we expect scanning SQUID
microscopy to have a lower sensitivity due to the increased distance
between the SQUID and the sample.

One can also tune the critical current value at which the transition
between laminar and turbulent regimes occurs by using materials with
different effective masses. For example, the heavy-hole effective
mass in p-doped GaAs is about $0.45m_e$, which implies that the
transition to turbulent flow in p-doped GaAs should occur at
$I'_{\rm c} = (0.067/0.45) *I_{\rm c} = 0.047 \, {\rm \mu A}$ for
the same doping density as the n-doped case.

This work was supported by the U.S.  Department of Energy under
grant DE-FG02-05ER46204.


\begin{thebibliography}{21}
\expandafter\ifx\csname natexlab\endcsname\relax\def\natexlab#1{#1}\fi
\expandafter\ifx\csname bibnamefont\endcsname\relax
  \def\bibnamefont#1{#1}\fi
\expandafter\ifx\csname bibfnamefont\endcsname\relax
  \def\bibfnamefont#1{#1}\fi
\expandafter\ifx\csname citenamefont\endcsname\relax
  \def\citenamefont#1{#1}\fi
\expandafter\ifx\csname url\endcsname\relax
  \def\url#1{\texttt{#1}}\fi
\expandafter\ifx\csname urlprefix\endcsname\relax\def\urlprefix{URL }\fi
\providecommand{\bibinfo}[2]{#2}
\providecommand{\eprint}[2][]{\url{#2}}

\bibitem[{\citenamefont{Madelung}(1926)}]{Madelung}
\bibinfo{author}{\bibfnamefont{E.}~\bibnamefont{Madelung}},
  \bibinfo{journal}{Z. Phys.} \textbf{\bibinfo{volume}{40}},
  \bibinfo{pages}{322} (\bibinfo{year}{1926}).

\bibitem[{\citenamefont{Bloch}(1933)}]{Bloch}
\bibinfo{author}{\bibfnamefont{F.}~\bibnamefont{Bloch}}, \bibinfo{journal}{Z.
  Phys.} \textbf{\bibinfo{volume}{81}}, \bibinfo{pages}{363}
  (\bibinfo{year}{1933}).

\bibitem[{\citenamefont{Martin and Schwinger}(1959)}]{MartinSchwinger}
\bibinfo{author}{\bibfnamefont{P.~C.} \bibnamefont{Martin}} \bibnamefont{and}
  \bibinfo{author}{\bibfnamefont{J.}~\bibnamefont{Schwinger}},
  \bibinfo{journal}{Phys. Rev.} \textbf{\bibinfo{volume}{115}},
  \bibinfo{pages}{1342} (\bibinfo{year}{1959}).

\bibitem[{\citenamefont{Vignale et~al.}(1997)\citenamefont{Vignale, Ullrich,
  and Conti}}]{vignale97}
\bibinfo{author}{\bibfnamefont{G.}~\bibnamefont{Vignale}},
  \bibinfo{author}{\bibfnamefont{C.~A.} \bibnamefont{Ullrich}},
  \bibnamefont{and} \bibinfo{author}{\bibfnamefont{S.}~\bibnamefont{Conti}},
  \bibinfo{journal}{Phys. Rev. Lett.} \textbf{\bibinfo{volume}{79}},
  \bibinfo{pages}{4878} (\bibinfo{year}{1997}).

\bibitem[{\citenamefont{Tokatly}(2005)}]{tokatly}
\bibinfo{author}{\bibfnamefont{I.~V.} \bibnamefont{Tokatly}},
  \bibinfo{journal}{Phys. Rev. B} \textbf{\bibinfo{volume}{71}},
  \bibinfo{pages}{165104} (\bibinfo{year}{2005}).

\bibitem[{\citenamefont{D'Agosta and {Di Ventra}}(2006)}]{dagosta06jpcm}
\bibinfo{author}{\bibfnamefont{R.}~\bibnamefont{D'Agosta}} \bibnamefont{and}
  \bibinfo{author}{\bibfnamefont{M.}~\bibnamefont{{Di Ventra}}},
  \bibinfo{journal}{J. Phys.: Condens. Matter} \textbf{\bibinfo{volume}{18}},
  \bibinfo{pages}{11059} (\bibinfo{year}{2006}).

\bibitem[{\citenamefont{Bushong et~al.}(2007)\citenamefont{Bushong, Gamble, and
  {Di Ventra}}}]{bushong07}
\bibinfo{author}{\bibfnamefont{N.}~\bibnamefont{Bushong}},
  \bibinfo{author}{\bibfnamefont{J.}~\bibnamefont{Gamble}}, \bibnamefont{and}
  \bibinfo{author}{\bibfnamefont{M.}~\bibnamefont{{Di Ventra}}},
  \bibinfo{journal}{Nano Lett.}  (\bibinfo{year}{2007}), \bibinfo{note}{in
  press}.

\bibitem[{\citenamefont{Topinka et~al.}(2000)\citenamefont{Topinka, LeRoy,
  Shaw, Heller, Westervelt, Maranowski, and Gossard}}]{topinka00}
\bibinfo{author}{\bibfnamefont{M.~A.} \bibnamefont{Topinka}},
  \bibinfo{author}{\bibfnamefont{B.~J.} \bibnamefont{LeRoy}},
  \bibinfo{author}{\bibfnamefont{S.~E.~J.} \bibnamefont{Shaw}},
  \bibinfo{author}{\bibfnamefont{E.~J.} \bibnamefont{Heller}},
  \bibinfo{author}{\bibfnamefont{R.~M.} \bibnamefont{Westervelt}},
  \bibinfo{author}{\bibfnamefont{K.~D.} \bibnamefont{Maranowski}},
  \bibnamefont{and} \bibinfo{author}{\bibfnamefont{A.~C.}
  \bibnamefont{Gossard}}, \bibinfo{journal}{Science}
  \textbf{\bibinfo{volume}{289}}, \bibinfo{pages}{2323} (\bibinfo{year}{2000}).

\bibitem[{\citenamefont{Chen and {Di Ventra}}(2003)}]{chediventra2003}
\bibinfo{author}{\bibfnamefont{Y.~C.} \bibnamefont{Chen}} \bibnamefont{and}
  \bibinfo{author}{\bibfnamefont{M.}~\bibnamefont{{Di Ventra}}},
  \bibinfo{journal}{Phys. Rev. B} \textbf{\bibinfo{volume}{67}},
  \bibinfo{pages}{153304} (\bibinfo{year}{2003}).

\bibitem[{\citenamefont{Gallop}(1991)}]{gallopsquids}
\bibinfo{author}{\bibfnamefont{J.~C.} \bibnamefont{Gallop}},
  \emph{\bibinfo{title}{SQUIDS, the Josephson effects and superconducting
  electroncs}} (\bibinfo{publisher}{IOP Publishing Ltd.},
  \bibinfo{year}{1991}).

\bibitem[{\citenamefont{Conti and Vignale}(1999)}]{conti99}
\bibinfo{author}{\bibfnamefont{S.}~\bibnamefont{Conti}} \bibnamefont{and}
  \bibinfo{author}{\bibfnamefont{G.}~\bibnamefont{Vignale}},
  \bibinfo{journal}{Phys. Rev. B} \textbf{\bibinfo{volume}{60}},
  \bibinfo{pages}{7966} (\bibinfo{year}{1999}).

\bibitem[{\citenamefont{D'Agosta et~al.}(2007)\citenamefont{D'Agosta, {Di
  Ventra}, and Vignale}}]{Robert}
\bibinfo{author}{\bibfnamefont{R.}~\bibnamefont{D'Agosta}},
  \bibinfo{author}{\bibfnamefont{M.}~\bibnamefont{{Di Ventra}}},
  \bibnamefont{and} \bibinfo{author}{\bibfnamefont{G.}~\bibnamefont{Vignale}},
  \bibinfo{journal}{Phys. Rev. B (in press)}  (\bibinfo{year}{2007}).

\bibitem[{\citenamefont{Popinet}(2003)}]{gerris}
\bibinfo{author}{\bibfnamefont{S.}~\bibnamefont{Popinet}}, \bibinfo{journal}{J.
  Comput. Phys.} \textbf{\bibinfo{volume}{190}}, \bibinfo{pages}{572}
  (\bibinfo{year}{2003}).

\bibitem[{\citenamefont{Lam and Tilbrook}(2003)}]{lam03}
\bibinfo{author}{\bibfnamefont{S.~K.~H.} \bibnamefont{Lam}} \bibnamefont{and}
  \bibinfo{author}{\bibfnamefont{D.~L.} \bibnamefont{Tilbrook}},
  \bibinfo{journal}{Appl. Phys. Lett.} \textbf{\bibinfo{volume}{82}},
  \bibinfo{pages}{1078} (\bibinfo{year}{2003}).

\bibitem[{\citenamefont{Ketchen et~al.}(1991)\citenamefont{Ketchen, Bhushan,
  Kaplan, and Gallagher}}]{noiseSQUID}
\bibinfo{author}{\bibfnamefont{M.~B.} \bibnamefont{Ketchen}},
  \bibinfo{author}{\bibfnamefont{M.}~\bibnamefont{Bhushan}},
  \bibinfo{author}{\bibfnamefont{S.~B.} \bibnamefont{Kaplan}},
  \bibnamefont{and} \bibinfo{author}{\bibfnamefont{W.~J.}
  \bibnamefont{Gallagher}}, \bibinfo{journal}{IEEE Trans. Magn.}
  \textbf{\bibinfo{volume}{MAG-27}}, \bibinfo{pages}{3005}
  (\bibinfo{year}{1991}).

\bibitem[{\citenamefont{Kirtley et~al.}(1995)\citenamefont{Kirtley, Ketchen,
  Tsuei, Sun, Gallagher, {Lock See Yu-Jahnes}, Gupta, Stawiasz, and
  Wind}}]{scanningSQUID}
\bibinfo{author}{\bibfnamefont{J.~R.} \bibnamefont{Kirtley}},
  \bibinfo{author}{\bibfnamefont{M.~B.} \bibnamefont{Ketchen}},
  \bibinfo{author}{\bibfnamefont{C.~C.} \bibnamefont{Tsuei}},
  \bibinfo{author}{\bibfnamefont{J.~Z.} \bibnamefont{Sun}},
  \bibinfo{author}{\bibfnamefont{W.~J.} \bibnamefont{Gallagher}},
  \bibinfo{author}{\bibnamefont{{Lock See Yu-Jahnes}}},
  \bibinfo{author}{\bibfnamefont{A.}~\bibnamefont{Gupta}},
  \bibinfo{author}{\bibfnamefont{K.~G.} \bibnamefont{Stawiasz}},
  \bibnamefont{and} \bibinfo{author}{\bibfnamefont{S.~J.} \bibnamefont{Wind}},
  \bibinfo{journal}{IBM J. Res. Dev.} \textbf{\bibinfo{volume}{39}},
  \bibinfo{pages}{655} (\bibinfo{year}{1995}).

\bibitem[{\citenamefont{Dinner et~al.}(2005)\citenamefont{Dinner, Beasley, and
  Moler}}]{Moler}
\bibinfo{author}{\bibfnamefont{R.~B.} \bibnamefont{Dinner}},
  \bibinfo{author}{\bibfnamefont{M.~R.} \bibnamefont{Beasley}},
  \bibnamefont{and} \bibinfo{author}{\bibfnamefont{K.~A.} \bibnamefont{Moler}},
  \bibinfo{journal}{Rev. Sci. Instrum. 76, 103702 (2005)}
  \textbf{\bibinfo{volume}{76}}, \bibinfo{pages}{103702}
  (\bibinfo{year}{2005}).

\bibitem[{\citenamefont{Veauvy et~al.}(2004)\citenamefont{Veauvy, Hasselbach,
  and Mailly}}]{veauvy04}
\bibinfo{author}{\bibfnamefont{C.}~\bibnamefont{Veauvy}},
  \bibinfo{author}{\bibfnamefont{K.}~\bibnamefont{Hasselbach}},
  \bibnamefont{and} \bibinfo{author}{\bibfnamefont{D.}~\bibnamefont{Mailly}},
  \bibinfo{journal}{Phys. Rev. B} \textbf{\bibinfo{volume}{70}},
  \bibinfo{pages}{214513} (\bibinfo{year}{2004}).

\bibitem[{\citenamefont{{Di Ventra} and Lang}(2002)}]{diventra02}
\bibinfo{author}{\bibfnamefont{M.}~\bibnamefont{{Di Ventra}}} \bibnamefont{and}
  \bibinfo{author}{\bibfnamefont{N.~D.} \bibnamefont{Lang}},
  \bibinfo{journal}{Phys. Rev. B} \textbf{\bibinfo{volume}{65}},
  \bibinfo{pages}{045402} (\bibinfo{year}{2002}).

\bibitem[{\citenamefont{Sai et~al.}(2007)\citenamefont{Sai, Bushong, Hatcher,
  and {Di Ventra}}}]{sai07}
\bibinfo{author}{\bibfnamefont{N.}~\bibnamefont{Sai}},
  \bibinfo{author}{\bibfnamefont{N.}~\bibnamefont{Bushong}},
  \bibinfo{author}{\bibfnamefont{R.}~\bibnamefont{Hatcher}}, \bibnamefont{and}
  \bibinfo{author}{\bibfnamefont{M.}~\bibnamefont{{Di Ventra}}},
  \bibinfo{journal}{Phys. Rev. B} \textbf{\bibinfo{volume}{75}},
  \bibinfo{pages}{115410} (\bibinfo{year}{2007}).

\bibitem[{\citenamefont{Liu et~al.}(2002)\citenamefont{Liu, Harris, and
  Kanwisher}}]{liu02}
\bibinfo{author}{\bibfnamefont{J.}~\bibnamefont{Liu}},
  \bibinfo{author}{\bibfnamefont{A.}~\bibnamefont{Harris}}, \bibnamefont{and}
  \bibinfo{author}{\bibfnamefont{N.}~\bibnamefont{Kanwisher}},
  \bibinfo{journal}{Nat. Neurosci.} \textbf{\bibinfo{volume}{5}},
  \bibinfo{pages}{910} (\bibinfo{year}{2002}).

\end{thebibliography}

\end{document}